\documentclass[a4paper]{jpconf}
\usepackage{graphicx}

\begin{document}
\rightline{\small IIT-CAPP-13-12}
\rightline{\small MICE-CONF-428}

\title{Progress Towards Completion of the MICE \leftline{Demonstration of Muon Ionization Cooling}}

\author{Daniel M. Kaplan}

\address{Illinois Institute of Technology, Chicago, IL 60616, USA}

\author{\vspace{
-.15in}(for the MICE Collaboration)}

\ead{kaplan@iit.edu}

\begin{abstract}
The Muon Ionization Cooling Experiment (MICE)  at the Rutherford Appleton Laboratory aims to demonstrate $\approx$\,10\%
ionization cooling of a muon beam by its interaction with low-$Z$ absorber materials followed by restoration of
longitudinal momentum in RF linacs.
MICE Step IV, including the first LH$_2$ or LiH absorber cell sandwiched between two particle tracking spectrometers, is the collaboration's near-term goal. Two  large
superconducting spectrometer solenoids and one focus coil
solenoid will provide a magnetic field of $\approx$\,4\,T in the
tracker and absorber-cell volumes. The status of these 
components is described, as well as progress towards Steps V and VI, including the eight RF cavities to provide the required 8\,MV/m gradient in a strong magnetic field; this entails an RF drive system to deliver 2\,MW, 1\,ms pulses of 201\,MHz frequency at 1\,Hz repetition rate,
the distribution network to deliver 1\,MW to each cavity with correct RF phasing, diagnostics to determine the gradient and the muon transit phase, and  development of the large-diameter magnets required for the linacs.\\[.1in]
(Presented at NuFact 2013, 15th International Workshop on Neutrino Factories, Super Beams and Beta Beams, 19--24 August 2013, IHEP, Beijing, China.)
\end{abstract}\vspace{-.2in}

\section{Introduction}
\begin{figure}
\centerline{\includegraphics[width=.62\linewidth, trim=450 0 300 0 mm,clip]{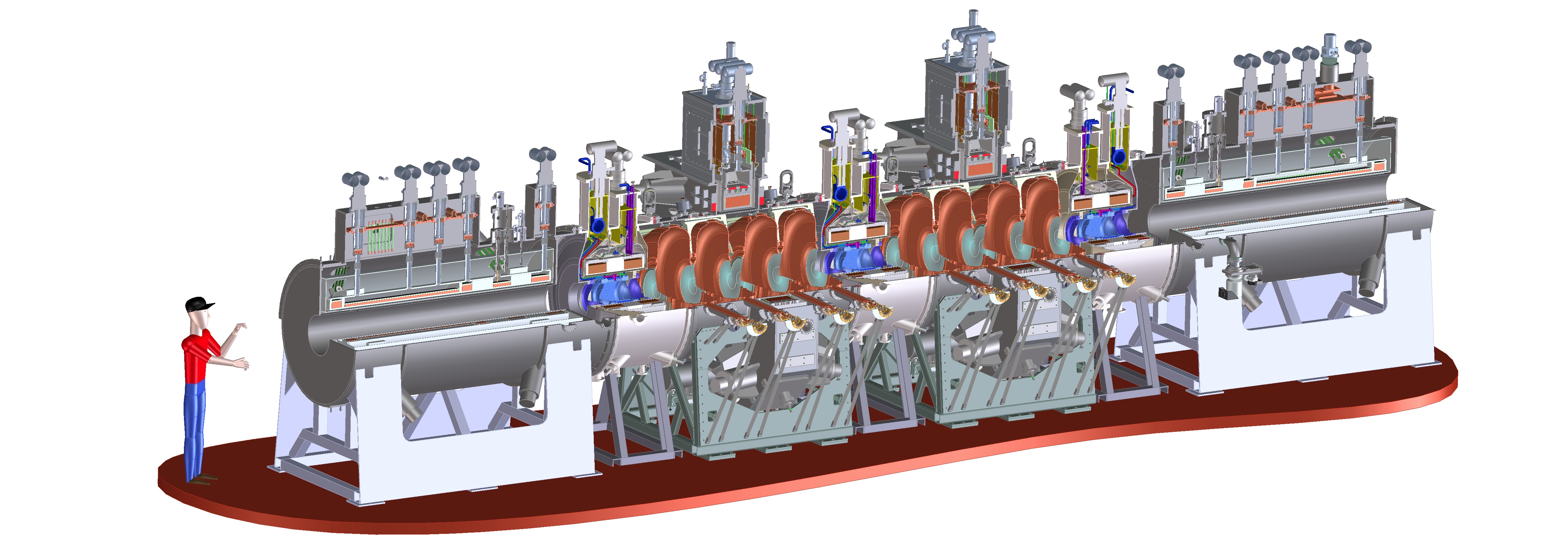}\hfill\includegraphics[width=.37\linewidth]{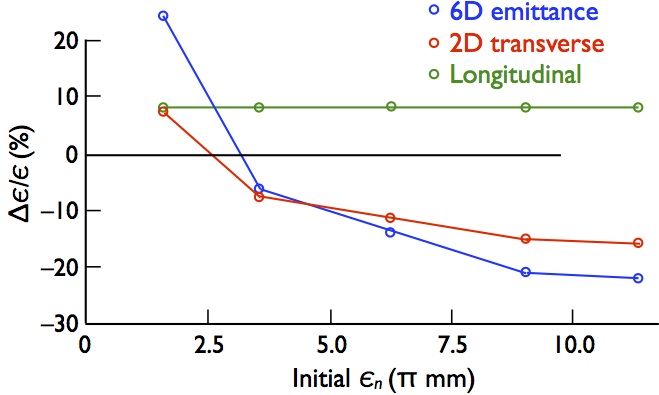}}
\vspace{-.1in}
\caption{Left: cutaway  rendering of  MICE (``Step VI") apparatus: two precision scintillating-fiber solenoidal spectrometers surround a cooling   cell based on the FS-II ``SFOFO" design~\cite{FSII} (particle-ID detectors not shown). Right: normalized-emittance change  vs input emittance in the nominal MICE optics configuration, showing 2.5$\pi$\,mm$\cdot$rad equilibrium  transverse emittance.\\[-.15in]
}\label{fig:MICE}
\end{figure}

The Muon Ionization Cooling Experiment~\cite{MICE} is under construction at the UK's Rutherford Appleton Laboratory (RAL) in order to demonstrate for the first time the feasibility and efficacy of ionization cooling of muons~\cite{cooling1,cooling2}. The MICE apparatus (Fig.~\ref{fig:MICE}) comprises one cooling lattice cell based on a design from Neutrino Factory Feasibility Study II (FS-II)~\cite{FSII} sandwiched by the input and output spectrometers and particle identification and timing detectors that will be used to  characterize the ionization-cooling process experimentally. Since an affordable cooling section cools by only $\sim$\,10\% (see Fig.~\ref{fig:MICE}), too small an effect to measure reliably using standard beam instrumentation,  MICE employs a low-intensity muon beam~\cite{beam} and measures each muon individually. Once completed, it will thereby demonstrate that the ionization-cooling process is  understood in detail in both its physics and engineering aspects, and that it works as simulated. In order to afford a thorough validation of the codes used to design ionization-cooling channels, MICE will be operated in various modes and optics configurations. Full results are expected by about 2020, with analyses of some configurations available up to five years earlier. Early results will include important validations of the models used in ionization-cooling simulation codes, as well as the first experimental test of muon transverse--longitudinal emittance exchange (needed for six-dimensional cooling, e.g., for a muon collider) in a wedge absorber.

Ionization cooling is inherently a transverse effect. A simple and intuitive way to see this is to consider that ionization energy loss exerts a braking effect on each muon, reducing both the transverse and longitudinal momentum components, while reacceleration in RF cavities restores only the longitudinal momentum component. With repeated braking and reacceleration the divergence of the beam is progressively reduced. Alternating-gradient focusing brings a concomitant reduction of the beam's cross-sectional area. The process can beneficially continue until an equilibrium is reached between the cooling effect of ionization and the heating effect of multiple scattering, i.e., until 
\begin{eqnarray} 
\frac{d\epsilon_n}{ds}\approx
-\frac{1}{\beta^2} \left\langle\!\frac{dE_{\mu}}{ds}\!\!\right\rangle\frac{\epsilon_n}{E_{\mu}}
 +
\frac{1}{\beta^3} \frac{\beta_\perp
(0.014\,{\rm GeV})^2}{2E_{\mu}m_{\mu}L_R}\to 0 \,,
\label{eq:cool} 
\end{eqnarray}  
 $d\epsilon_n/ds$ being the rate of normalized-emittance change within the absorber; $\beta c$,  $E_\mu$,  and $m_\mu$
the muon velocity, energy, and mass; $\beta_\perp$ the lattice betatron function at the absorber; and $L_R$ the  radiation length of the absorber material~\cite{cooling2}. Ionization cooling works optimally for momenta near the ionization minimum, hence MICE is designed for the momentum range 140 to 240\,MeV/$c$.

\section{MICE Steps}

\begin{figure}
\begin{minipage}[b]{20pc}
\includegraphics[width=1.13\linewidth
]{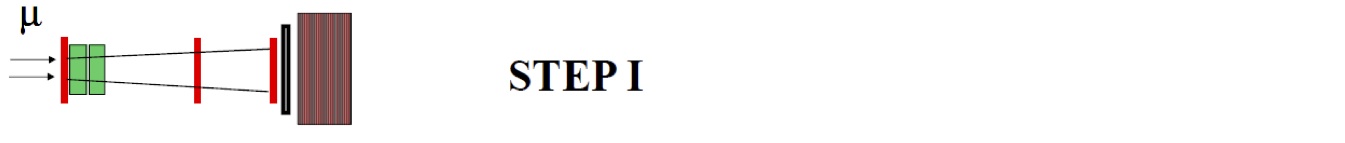}\\
\includegraphics[width=1.03\linewidth,
]{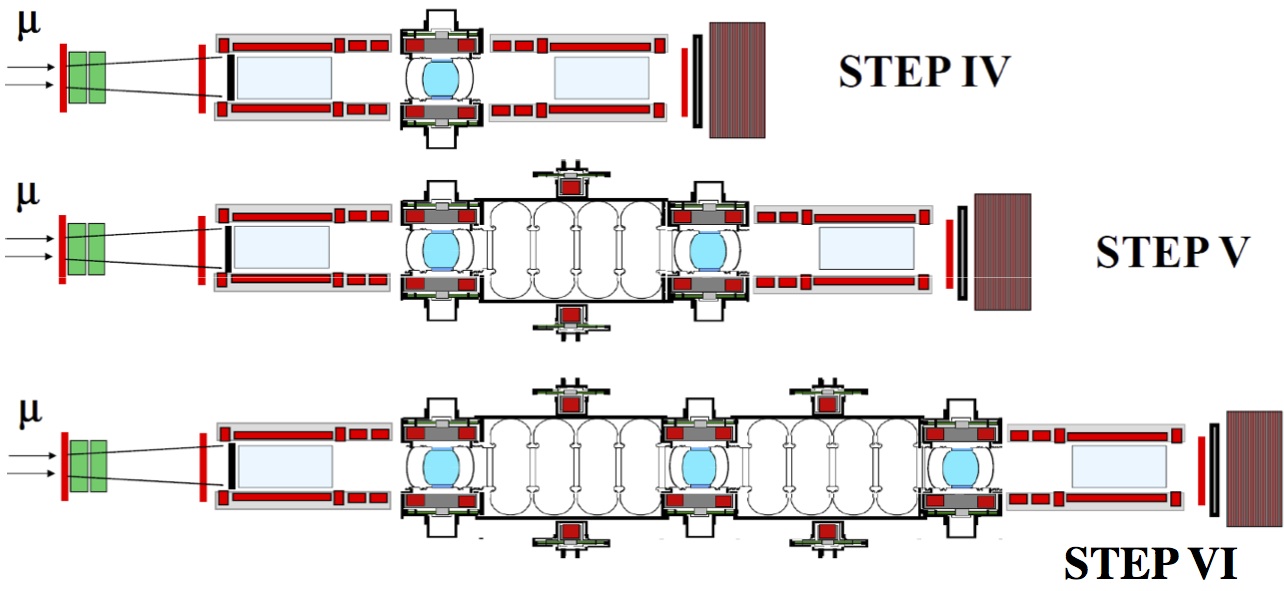}
\caption{The Steps of MICE}\label{fig:Steps}
\end{minipage}\hfill
\begin{minipage}[b]{16pc}
\centerline{\includegraphics[width=\linewidth]{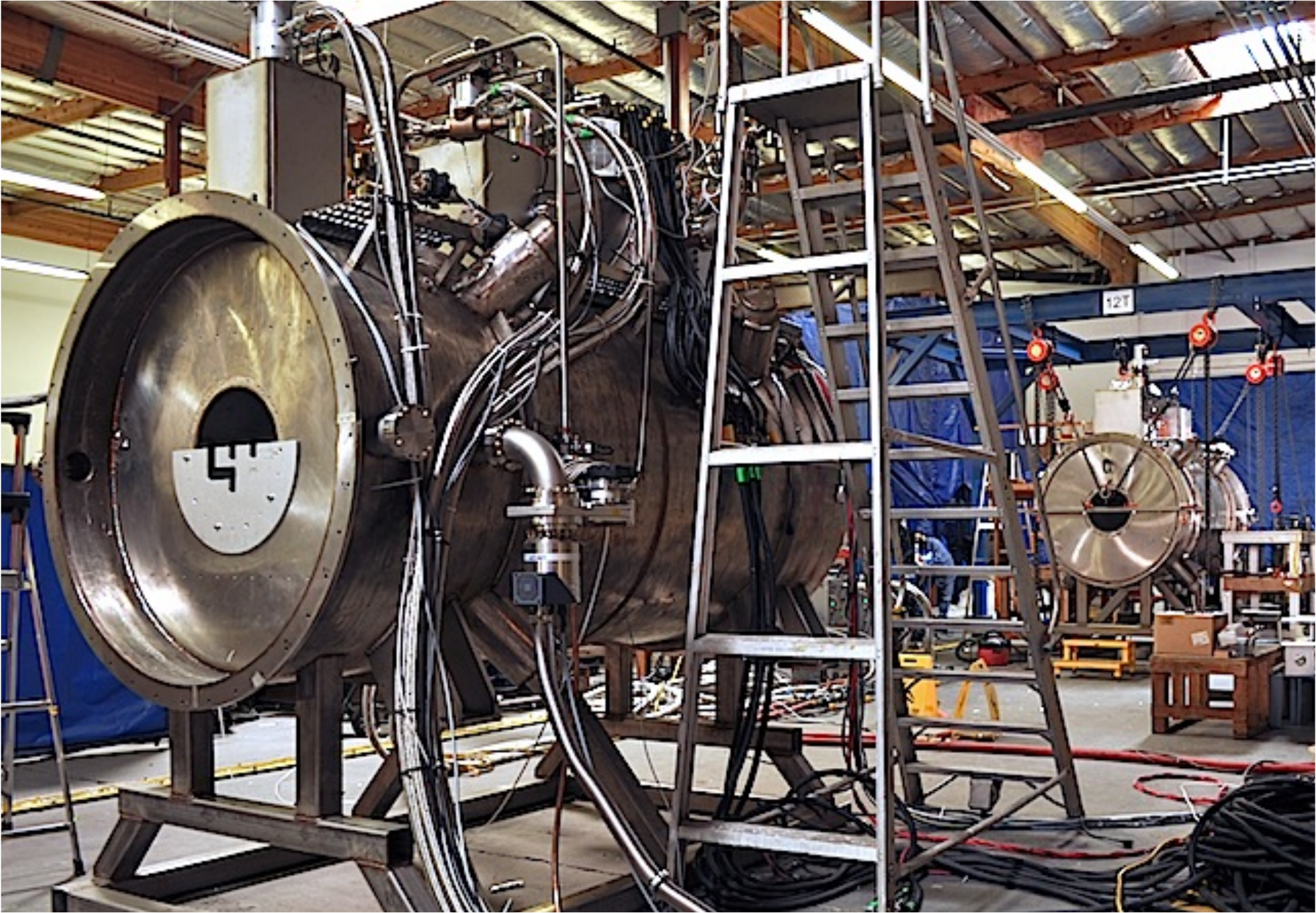}}
\caption{MICE spectrometer solenoids at vendor}\label{fig:SS}
\end{minipage}

\end{figure}

The MICE precision goal, 0.1\% emittance resolution, allows even the small emittance decrements (increments) when operating just above (below) the cooling cell's equilibrium emittance to be well measured. This  places a premium on careful spectrometer characterization and calibration. This goal and the anticipated funding schedule led to a planned MICE buildup in a series of six ``Steps," but  exigencies of complex apparatus construction have resulted in the abbreviated sequence of Fig.~\ref{fig:Steps}. Step I has already taken place, and the results presented~\cite{beam,PID}, with the exception of the  Oct.\ 2013 EMR (``electron--muon ranger" total-absorption calorimeter) run. 

\section{Step IV}

MICE Step IV will represent a first study of muon ionization cooling, with results expected starting in 2015. 
In addition to the Step I components (pion-production target inserted periodically into the ISIS synchrotron beam, pion-to-muon beamline,  particle-ID and muon-timing detectors), Step IV requires the two trackers with their spectrometer solenoids (SS) and one absorber--focus-coil (AFC) module. The trackers are in hand and have been tested with cosmic rays~\cite{tracker}. Each SS comprises five superconducting coils: three to provide a 4\,T field uniform to better than 1\% over the 1-m-long, 40-cm-diameter tracking volume, and two to match the beam into or out of the  cooling cell. The  spectrometer solenoids (Fig.~\ref{fig:SS}) have been built by Wang NMR in Livermore, CA, USA, and the focus coils by Tesla Engineering Ltd., UK. As of this writing, after some retrofitting, the first SS has been successfully trained, field-mapped, and delivered to RAL, and training of the second has commenced. The first  AFC has been trained to just above the baseline current in ``flip" mode (the two coils powered in opposite polarities) and to the full design current in ``solenoid" mode (same polarities). The training of a second AFC is commencing and the performance of the first will be assessed in the light of that experience.

Step IV will include tests of LH$_2$, LiH, and plastic or other low-$Z$ absorbers. A demonstration of emittance exchange is also planned. Muon colliders require muon cooling in all six phase-space dimensions, and rely on emittance exchange in order to couple the transverse  ionization cooling effect into the longitudinal phase plane. This is accomplished via a suitable correlation between muon momentum and path length in absorbers, and thus requires lattices with dispersion. While there is naturally some dispersion in the MICE beamline, since in MICE each muon is measured individually,  the desired range of dispersion will be created via off-line selection of muons.

\section{Toward Step VI}

MICE Step IV will allow the detailed characterization of ionization energy loss, multiple Coulomb scattering, and their effect on the ionization cooling process, and will thus validate the models used in muon ionization-cooling simulations. However, it constitutes a ``non-sustainable" cooling configuration, since the energy lost in the absorber is not replenished.  The first engineering demonstration of  ``sustainable" cooling will thus be  Step VI (depicted in Fig.~\ref{fig:MICE}), with a possible ``stopover" in Step V, depending on component availability. By allowing a thorough exploration of the optics of ionization-cooling lattices, including those with periodic field flips  and those with a modulated solenoid field, Step VI will furnish a complete validation of the simulations.

\begin{figure}
\begin{minipage}[b]{17pc}
\centerline{\includegraphics[width=\linewidth,trim=40 0 40 0 mm,clip]{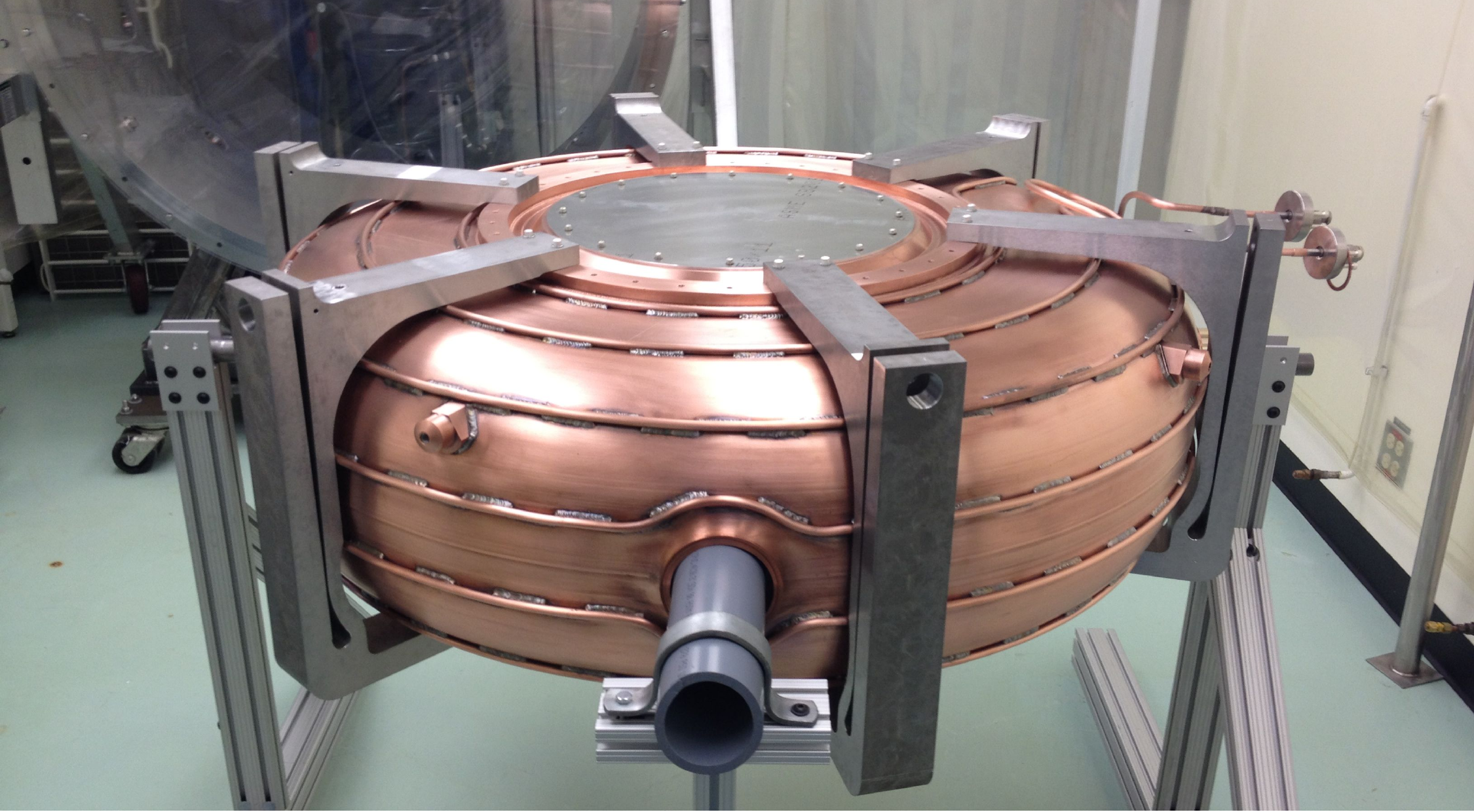}}
\caption{First MICE RF cavity being fitted with tuners at Fermilab.}\label{fig:201MHz}
\end{minipage}\hfill
\begin{minipage}[b]{17pc}
\centerline{\includegraphics[width=\linewidth,trim=0 0 0 10 0mm,clip]{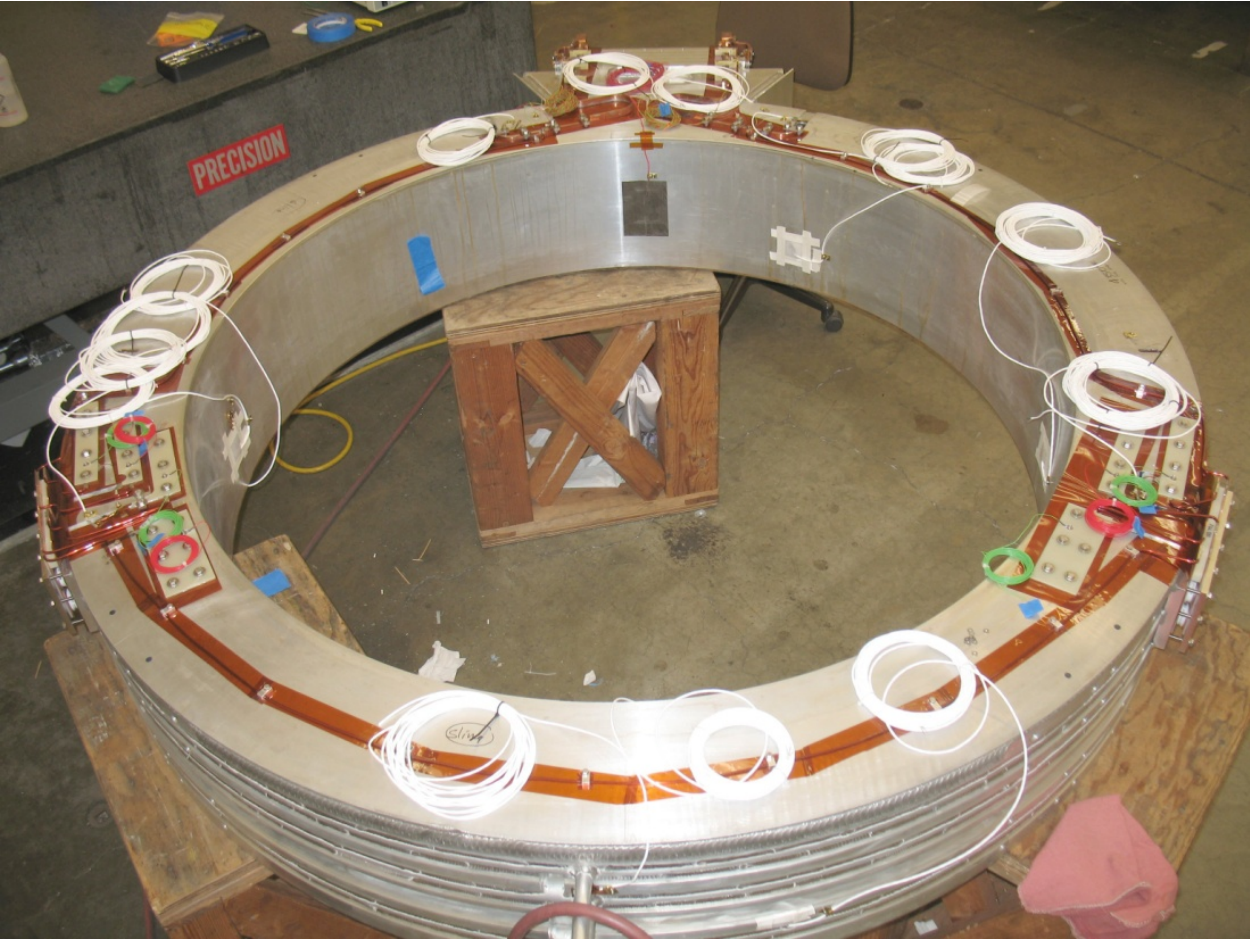}}\vspace{-.1in}
\caption{First CC cold mass fitted out at LBNL prior to shipment to Fermilab.}\label{fig:CC}
\end{minipage}
\end{figure}

Step VI requires (besides the components of Step IV) two additional AFC modules and two RF-cavity--coupling-coil (RFCC) modules, each containing four 201\,MHz RF cavities and one large-diameter coupling-coil (CC) solenoid. All eight RF cavities have been fabricated and the first (Fig.~\ref{fig:201MHz}) is being readied for testing in the Fermilab MuCool Test Area (MTA). At the FS-II design gradient these cavities each require 4 MW of input power, provided through two coaxial couplers. Prototype tests  at the MTA revealed RF-coupler breakdown as an issue. Couplers of a revised design are now in fabrication and will soon be tested in the MTA. 

A full test requires in addition a coupling coil, so that the behavior of the cavity in a multi-tesla solenoid field can be explored. The first CC cold mass (Fig.~\ref{fig:CC}), built for Harbin Institute of Technology by Qi Huan Co., Beijing, China, and planned for the MTA rather than MICE, is now under test at Fermilab. Once it is trained, three more (including one spare) will be fabricated. Fabrication of the needed cryostats and vacuum vessels 
{(at LBNL, based on initial designs developed by SINAP)}, and of the cavity tuners, is also in progress, as is assembly of the RF control and power distribution systems by LBNL and Daresbury Lab.

The FS-II design~\cite{FSII} calls for 32\,MW of RF power per cell in order to allow off-crest operation. The MICE Step VI goal is on-crest acceleration, with power provided by four 2\,MW supplies  from CERN and LBNL refurbished by Daresbury Lab. The first supply has recently reached full power, with a single-supply test  at MICE scheduled for later this year. 

\section{Conclusions}
MICE is progressing towards the first experimental study of muon ionization cooling. Step IV is planned for 2015 and the concluding Step VI  of MICE by the end of this decade. Once ionization cooling is established, construction of a neutrino factory with cooling could commence. 

\section*{Acknowledgments}
The author thanks his MICE collaborators 
for many stimulating interchanges on these topics. Work supported by the U.S.\ Dept.\ of Energy (through the Muon Accelerator Program) and National Science Foundation.

\end{document}